\documentclass[12pt,twoside]{article}
\usepackage{psfig}
\newcommand{\VolumeHeader}{}
\newcommand{\VolumeSerial}{LNS}

\newcommand{\ActivityName}{ {\normalsize {\it 
Conference on Gravitational Waves:
A Challenge to Theoretical Astrophysics
}}}
\newcommand{\ActivityDate}{ {\normalsize {\it
Trieste, 5-9 June 2000
}}}

\newcommand{\LectureHeader}{Quasi-Stationary Binary Inspiral}

\begin{document}
\pagestyle{myheadings}
\markboth{\LectureHeader}{\VolumeHeader}
\markright{\VolumeHeader}


\begin{titlepage}


\title{A Stationary Approximation to the Spacetime\\
of a Compact Object Binary} 

\author{J.T.Whelan$^{\dagger\ddagger}$\thanks{jtwhelan@utb1.utb.edu}
  \\[1cm]
  {\normalsize {\it $^\dagger$ Department of Physical Sciences, The
      University of Texas at Brownsville,}}
  \\
  {\normalsize {\it Brownsville, TX, USA.}}
  \\
  {\normalsize {\it $^\ddagger$ Institut f\"{u}r theoretische Physik,
      Universit\"{a}t Bern, Bern, SCHWEIZ}}
  \\[10cm]
{\normalsize {\it Lecture given at the: }}
\\
\ActivityName 
\\
\ActivityDate 
\\[1cm]
{\small \VolumeSerial} 
}
\date{}
\maketitle
\thispagestyle{empty}
\end{titlepage}

\baselineskip=14pt
\newpage
\thispagestyle{empty}


\begin{abstract}

  The gravitational-radiation-induced inspiral of a binary system of
  compact objects is considered.  A scheme is described to model the
  regime in which the gravitational interaction is too strong to use
  weak-field approximation methods, but the time scale for decay of
  the orbits is still long compared to the orbital period, by
  numerically solving for a stationary spacetime which approximates
  the slowly evolving one.  Equilibrium is to be maintained in the
  radiating system by imposing a balance of incoming and outgoing
  radiation at large distances.  Numerical results from non-linear
  scalar field theory have shown that such an approach can be
  effective modelling a slowly evolving solution to a wave equation.

\end{abstract}

\vspace{6cm}




\newpage
\thispagestyle{empty}
\tableofcontents

\newpage
\setcounter{page}{1}

\section{Introduction}

A pair of compact objects (black holes or neutron stars) in binary
orbit will, according to Newtonian gravity, remain in the same orbits
forever.  In general relativity, however, the system will emit
gravitational radiation, causing the bodies to spiral in towards one
another.  The gravitational radiation given off by this system is a
prime candidate for detection by upcoming gravitational wave
telescopes such as VIRGO and LIGO \cite{Rowa00}.

Full three-plus-one numerical evolution of the Einstein equations is a
powerful tool for determining the history of such a system and the
form of the gravitational radiation emitted, but is limited in its
applicability by instabilities which prevent the simulation from being
evolved for times longer than several orbital periods. Thus it is not
only efficient but also necessary to limit full-numerical simulations
to the portion of the evolution for which there is not an applicable
approximation scheme \cite{Bake00}. The early stages of inspiral can be
handled by post-Newtonian methods \cite{Buon00}, while the final
post-merger ringdown can be treated with black hole perturbation
theory \cite{Pull98}.  The purpose of the quasi-stationary
approximation is to provide improved early waveforms and later-time
supercomputer initial conditions by modelling a late inspiral phase,
in which some non-perturbative gravitational effects are relevant, but
the radiation-induced inspiral is still occurring slowly.

\section{General Formulation}

\subsection{Quasi-Stationary Approximation}\label{ssec:qsapprox}
  
The idea, initially proposed by Steven Detweiler  \cite{Blac92,Detw94},
is that if the inspiral is slow, the system is nearly periodic: after
one orbit, the objects have returned almost to their original
locations, and radiation which has moved out has been replaced with
new radiation of approximately the same shape.  If the objects' orbits
are circular rather than elliptical, the spacetime is nearly
stationary.  If the approximate orbital frequency is $\Omega$, moving
forward in time by $\delta t$ and rotating the resulting spatial slice
by $-\Omega\,\delta t$ will not change the picture very much.

This approximation can be used to simplify the numerical problem by
solving for an exactly stationary spacetime which serves, over some
period of time, as a reasonable approximation for the slowly evolving
spacetime.  In the process, the three-plus-one dimensional evolution
problem is reduced to a three-dimensional, non-dynamical one, not only
reducing the size of the computational grid, but also hopefully
avoiding evolutionary instabilities.

Since gravitational radiation emitted from the orbiting system carries
away energy, some modification must be made to the physical problem in
order to allow this equilibrium solution.  One approach
\cite{Cook96,Wils95,Wils96} is to require the spatial geometry of the
stationary spacetime to be conformally flat and solve for the
conformal factor using only a subset (the constraints) of the Einstein
equations.  This method enforces stationarity by in effect discarding
degrees of freedom associated with the radiation which would carry
away energy.  Our approach, on the other hand, is to keep the
gravitational radiation, but nullify the radiation reaction by
balancing the outgoing radiation by an equal amount of incoming
radiation.  In so doing, we will solve the full Einstein equations in
the presence of the actual sources, and simply replace the physical
boundary condition of outgoing radiation at large distances with one
of balanced radiation.

\subsection{Radiation-Balanced Boundary Conditions}\label{ssec:rbbc}

In the general context of a theory in which a pair of orbiting sources
for a wave equation give off radiation, it is instructive to consider
three types of solutions.  In a \textbf{Type~I} solution, the
radiation is outgoing, carrying away energy and causing the orbits to
decay.  In a \textbf{Type~II} solution, the radiation is again
outgoing, but some additional force, uncoupled to the radiating field,
keeps the sources in their circular orbits, resulting in a stationary
spacetime.  A \textbf{Type~III} solution is also stationary, with the
sources remaining in circular orbits of a constant frequency, but this
time there is no external force, and the equilibrium is maintained by
a balance of incoming and outgoing radiation.  Type~I is the physical
situation we ultimately wish to model.  Type~II should be a reasonable
approximation to Type~I if the Type~I solution in question is
inspiralling only slowly.  It should not be appropriate to general
relativity, in which all matter and energy couples to the
gravitational field, but is useful for comparison in theories which
allow for external forces.  Type~III is the stationary solution we
wish to find numerically, and, where appropriate, to relate it to a
Type~II or I solution.

To analyze in detail the implications of boundary conditions on
radiating, stationary solutions (Types II and III), we have considered
the theory of a non-linear scalar field $\psi(t,\rho,\phi)$ in a $2+1$
dimensional Minkowski spacetime, the simplest theory with
nonlinearity, orbits, and radiation.  If the source is required to
rotate at a constant angular velocity $\Omega$, so that the charge
density $\sigma(t,\rho,\phi)$ is a function only of $\rho$ and
$\varphi=\phi-\Omega t$, there exist stationary solutions for which
the field exhibits the same symmetry; in that case, the wave equation
can be written
\begin{equation}
  \label{eq:corot}
  \frac{\partial^2\psi}{\partial\rho^2}
  +\rho^{-1}\frac{\partial\psi}{\partial\rho}
  +\left(
    \rho^{-2}-\Omega^2 c^{-2}
  \right)    
  \frac{\partial^2\psi}{\partial\varphi^2}
  = -\sigma + \lambda \mathcal{F}(\psi,\rho) 
  ,
\end{equation}
where $\mathcal{F}(\psi,\rho)$ is a non-linear coupling.  A convenient
choice of charge distribution $\sigma$ is a positive charge at a
radius $a$ and azimuthal angle $\varphi=0$ and an equal and opposite
charge at $\varphi=\pi$ and the same radius.  This distribution
satisfies $\sigma(\rho,-\varphi)=\sigma(\rho,\varphi)$.

In the context of a numerical solution on a finite grid, the Type~II
(purely outgoing radiation) solution $\psi^R_{\mathrm{out}}$ is
defined by applying a Sommerfeld ($\psi_{,\rho}+c^{-1}\psi_{,t} =
\psi_{,\rho}-\Omega c^{-1}\psi_{,\varphi}=0$) boundary condition at a
the radius $R$ of the outer boundary of the grid.  If the non-linear
term becomes small at large distances, the form of any solution near a
large-$R$ boundary will be a solution to the vacuum, linear version of
(\ref{eq:corot}); each angular Fourier mode will be a linear
combination of two independent solutions, which can be identified as a
purely ingoing and a purely outgoing solution.  The limit
$\psi_{\mathrm{out}} := \lim_{R\rightarrow\infty}
\psi^R_{\mathrm{out}}$ is simply the solution consisting only of
outgoing modes, with the co\"{e}fficients of all the ingoing modes set
to zero.  [The entire discussion can be repeated for ingoing
radiation, with the resulting solution being
$\psi_{\mathrm{in}}(\rho,\varphi)=\psi_{\mathrm{out}}(\rho,-\varphi)$.]

The most familiar Type~III radiation-balanced (RB) solutions are
standing-wave solutions in which the Dirichlet ($\psi=0$) or Neumann
($\psi_{,\rho}=0$) boundary condition is enforced at $\rho=R$.  If the
large-distance form of any of these solutions is resolved in angular
Fourier modes, it is found that the amplitudes of the co\"{e}fficients
of the ingoing and outgoing components are equal; however the relative
phase of the two independent vacuum solutions depends on the choice of
$R$, as does the radiation amplitude corresponding to a given source
strength.  Unlike outgoing- (or ingoing-) wave solutions,
standing-wave solutions do not tend towards any limit as the boundary
radius is taken to infinity.

Standing waves are not the only solutions with equal magnitudes of
incoming and outgoing radiation; in the case of a linear theory
($\lambda=0$) another such solution is simply a linear superposition
of the ingoing and outgoing solutions (LSIO) $\psi_{\mathrm{LSIO}}
=\frac{1}{2}(\psi_{\mathrm{in}}+\psi_{\mathrm{out}})$.  For a given
source strength, this turns out to be the RB solution with the
smallest amplitude, which means in effect that it's the solution with
just the radiation needed to keep the sources in equilibrium and no
``extra'' radiation.  From the point of view of using a Type~III
solution to approximate a Type~II one, we can take this ``minimum
energy radiation balance'' (MERB) solution, extract its outgoing
component, and identify it with the outgoing radiation.

In the non-linear theory, the superposition of two solutions is no
longer a solution, but the same effect can be achieved by
reformulating (\ref{eq:corot}) as a Green's function problem and using
the average of the advanced and retarded Green's functions to find a
MERB solution.  We have done this \cite{Whel00} and found that even
for highly non-linear theories, this time-symmetric Green's function
method can be used to find a good approximation for the outgoing
solution.  This lends confidence that in the case of general
relativity, if we can find a Type~III solution with little or no
``superfluous'' radiation, we may be able to relate it to the physical
Type~I solution in an analogous way.

As an aside, it is worth stressing that while the equation
(\ref{eq:corot}) is elliptical inside and hyperbolic outside the
``speed of light circle'' $\rho=1/\Omega$, we experienced no
difficulties finding the numerical solution to the problem using
closed-surface boundary conditions usually associated with a purely
elliptical equation.  In particular there were no discernable
artefacts at the light circle, which required no special treatment in
the numerical solution.

\section{Cosmic String Toy Model}

To apply the quasi-stationary method to general relativity, we first
note that the stationarity of the numerical spacetime is described by
a Killing vector $K_0\sim\partial_t+\Omega\partial_\phi$.  This means
that the geometry of the four-dimensional spacetime is determined by
the geometry on a representative three-dimensional surface.  Before
attacking this three-dimensional problem, we first consider a
computationally simpler toy model in which the sources and fields
posess an additional Killing symmetry $K_1\sim\partial_z$
corresponding to a translation perpendicular to the orbital plane.
This then is a model describing not orbiting localized sources, but
rather orbiting infinitely long lines of mass, or cosmic
strings.\footnote{The cosmic strings in question posess a curvature
  singularity along the string and generate Riemann curvature in the
  external spacetime.  They are not to be confused with the more
  familiar special case of cosmic strings consisting of a conical
  singularity surrounded by a flat external spacetime.}  The addition
of a second Killing vector field in four-dimensional spacetime means
that the geometry need only be numerically determined on a
two-dimensional surface.  Our primary intention in looking at this
model is to test the method rather than to obtain astrophysically
meaningful results, but it's worth pointing out that this would
provide a solution to the two-body problem in full vacuum general
relativity, albeit with somewhat unusual boundary conditions.

While numerically simpler, the problem of orbiting cosmic strings
brings with it some conceptual subtleties, which are described in the
next two subsections.

\subsection{Spacetimes with Two Killing Vectors}

One question is how to choose a basis which takes full advantage of
the two Killing symmetries of the stationary orbiting cosmic string
spacetime, thereby removing any gauge freedom from the numerical
problem.  Since the orbits of $K_0$ (which is associated with the
stationarity of the spacetime and also present in the
three-dimensional case) are not surface forming, a co\"{o}rdinate
basis on the spacetime four-manifold turns out not to be ideally
suited to the task.  Instead, one defines co\"{o}rdinates $x^2$ and
$x^3$ on the two-manifold of Killing-vector orbits (every point of
which represents an equivalence class of points in the four-manifold
connected by Killing trajectories.  The associated basis vectors $e_2$
and $e_3$ commute when restricted to the two-manifold, but the
corresponding vectors they define on the four-manifold do not commute.
Thus $e_2$ and $e_3$, together with the Killing vectors $K_0$ and
$K_1$, define a non-co\"{o}rdinate basis on the spacetime
four-manifold.

Extending the results of of \cite{Gero72}, we have used this method to
find explicit forms of the Einstein equations, both in a general
two-Killing-vector spacetime and in the case of orbiting cosmic
strings \cite{Whel99}.  The most convenient choice of co\"{o}rdinates
($\rho,\varphi$) on the two-manifold of Killing-vector orbits are the
anologue of the corotating polar co\"{o}rdinates used in
Section~\ref{ssec:rbbc}.

\subsection{Perturbations to Levi-Civita}

To analyze and apply boundary conditions to the radiation far from the
sources, we need to cast the spacetime in a form consisting of a
static background plus a perturbation describing gravitational
radiation on that background.  With localized sources, that background
could in principle be taken to be Minkowski spacetime, but this is
impossible with line-like sources because the spacetime of an infinite
line of mass is not asymptotically flat, even as one approaches
infinity in directions orthogonal to the line.  (Equivalent ways of
seeing this are to note that the gravitational potential of a line
mass falls off logarithmically, or that the quantity
mass-per-unit-length is dimensionless in general relativity.)  The
simplest appropriate background spacetime is the static, cylindrically
symmetric Levi-Civita spacetime \cite{Levi19} describing a single
cosmic string; the Riemann curvature is determined by a single
dimensionless parameter $C$ which is associated with the mass per unit
length of the string.

Analysis of perturbations to Levi-Civita \cite{Whel01} shows that
tensor as well as scalar modes are governed by an equation
\begin{equation}\label{eq:lcmode}
 \partial_\rho^2 \psi_m(\rho) + \rho^{-1} \partial_\rho \psi_m(\rho)
 + F_m(\rho) \psi_m(\rho) = 0
 .
\end{equation}
The potential $F_m(\rho)$ depends on the tensor character and
parity of the mode considered,  but the leading
term at large $\rho$ is
\begin{equation}
 F(\rho) \sim m^2 \Omega^2 \rho^{-2C/(1-C+C^2)}
 .
\end{equation}
[Note that for the flat background case $C=0$, this reduces to the
large-$\rho$ limit of the linear version of (\ref{eq:corot}).]

\subsection{Strategy for Radiation Balance in General Relativity}

The general solution to (\ref{eq:lcmode}) is, as before, a linear
combination of an outgoing and an ingoing solution.  In principle, a
general RB solution can be defined as one with equal ingoing and
outgoing amplitudes.  However, for the purpose of numerically
enforcing RB, it is simpler to use an equivalent condition.  If the
locations of the sources are taken to be at $\varphi=0$ and
$\varphi=\pi$, the condition of RB is equivalent to invariance of the
geometry under the reflection $\varphi\rightarrow -\varphi$.  This can
be seen by noting that the flux $T_{t\rho}$ of energy through a
surface of constant $\rho$ will be proportional to
$\psi_{,\rho}\psi_{,\varphi}$.  If $\psi$ is even under
$\varphi$-inversion, $\psi_{,\rho}$ will be even and $\psi_{,\varphi}$
odd, making $T_{t\rho}$ odd under $\varphi\rightarrow -\varphi$.  This
means that the total rate of energy (integrated over $\phi$ or
equivalently $\varphi$) flowing through a constant-$\rho$ surface will
vanish for a field configuration obeying this reflection symmetry.  If
the sources are taken to have equal mass-per-unit-length, the geometry
will also be unchanged by a $180^\circ$ rotation which interchanges
them, meaning that one need only solve for the metric over the
interval $\varphi\in[0,\pi/2]$.  This ``quadrant symmetry'' is the
analogue in the two-dimensional problem of the octant symmetry familiar
from \cite{Cook96,Wils95,Wils96}; it is not only a computational
simplification but also a means of automatically ensuring a balance of
radiation.

As noted in Section~\ref{ssec:rbbc}, a wave equation has not one but
an entire family of RB (Type~III) solutions.  We would like to choose
the one which can be most directly related to the (non-stationary)
Type~I solution with outgoing radiation.  While the nature of the full
nonlinear Einstein equations means that the Green's function method
used for scalar field theory will not work in the gravitational case,
we can still try to select a RB solution with as little gravitational
radiation as necessary, applying the following scheme:
\begin{enumerate}
\item Treating the sources perturbatively, obtain a guess for the
  relative phase of the ingoing and outgoing components of the
  radiation and thus a candidate asymptotic form for each mode of the
  MERB;
\item Apply a boundary condition to the values of the metric
  components at the outermost two $\rho$ values in the grid which
  forces the perturbation to be proportional to the candidate MERB;
\item Modify the candidate MERB and repeat step 2 to get a family of
  solutions, and find the one with the smallest total gravitational
  wave amplititude.
\end{enumerate}

\section{Conclusions and Outlook}

This paper has described an ongoing research program to find
numerically a stationary spacetime which can approximate over some
stretch of time a slowly inspiralling compact object binary.  A
stationary solution to the radiative problem is to be achieved by
replacing the physical boundary condition of purely outgoing radiation
with a condition corresponding to an equal mix of ingoing and outgoing
radiation.

Thus far, the numerical work has been limited to testing the
usefulness of this sort of radiation-balanced approach in non-linear
scalar field theory, but we have formulated a toy problem in general
relativity with infinitely extended sources and are about to begin
analyzing it numerically.  Looking beyond, to the application of the
method to the physical problem of localized sources, it is instructive
to consider how such an equilibrium three-dimensional spacetime could
be used.

First, since the spacetime will contain at large distances a
superposition of ingoing and outgoing gravitational radiation, we
should be able to separate out the outgoing-wave contribution and use
it as an approximation for the purely-outgoing radiation from the
physical system.

Second, a section of the spacetime could be used as an alternative to
post-Newtonian or conformally flat initial data for a full numerical
evolution.  (Since our approach contains in some sense ``too much''
and the conformally flat approach ``too little'' radiation, a
comparison of the two should provide insight into the effects of the
radiation on the evolution of the spacetime.)  It should be pointed
out that while the RB spacetime is an equilibrium solution of
Einstein's equations, valid for all time, it can still be used to
provide initial data for a full numerical evolution, since the outer
boundary condition of such an evolution will be one of outgoing rather
than balanced radiation, and thus once the ``missing'' ingoing
radiation has failed to propagate from the outer boundary to the
location of the sources, the orbits will begin to decay.  (The author
is indebted to B.~Br\"{u}gmann for clarifying this point.)

\section*{Acknowledgments}

In addition to my collaborators R.~Price, J.~Romano and W.~Krivan, I
would like to thank S.~Detweiler, P.~Brady, T.~Creighton,
\'{E}.~Flanagan, S~.Hughes, K.~Thorne, A.~Wiseman, C.~Torre,
J.~Friedmann, S.~Morsink, A.~Held, P.~H\'{a}j\'\i\v{c}ek,
J.~Bi\v{c}\'{a}k, J.~Friedman, J.~Novak, J.~Baker, B.~Br\"{u}gmann,
M.~Campanelli, and C.~Lousto.  This work was partially supported by
the National Science Foundation under grant PHY9734871, by the Swiss
Nationalfonds, and by the Tomalla Foundation, Z\"{u}rich.


\end{document}